\documentclass[prb,aps,twocolumn,showpacs,superscriptaddress,floatfix]{revtex4-1}
\usepackage{amsmath}
\usepackage{amssymb}
\usepackage{amsfonts}
\usepackage{mathptmx}

\usepackage{graphicx}
\usepackage{epstopdf}
\usepackage{dcolumn}
\usepackage{bm}

\DeclareMathOperator{\tr}{\mathop{\mathrm{Tr}}}

\DeclareMathOperator{\re}{\mathop{\mathrm{Re}}}
\DeclareMathOperator{\im}{\mathop{\mathrm{Im}}}

\newcommand{\Eq}[1]{Eq.~(\ref{#1})}
\newcommand{\Eqs}[1]{Eqs.~(\ref{#1})}

\begin{document}

\title{Electron cooling in  diffusive normal metal - superconductor tunnel junctions with a spin-valve ferromagnetic interlayer}
\date{\today}
\author{A.~Ozaeta}
\affiliation{Centro de F\'{i}sica de Materiales (CFM-MPC), Centro
Mixto CSIC-UPV/EHU, Manuel de Lardizabal 5, E-20018 San
Sebasti\'{a}n, Spain}
\author{A.~S.~Vasenko}
\affiliation{Institut Laue-Langevin, 6 rue Jules Horowitz, BP 156,
38042 Grenoble, France}
\author{F.~W.~J.~Hekking}
\affiliation{LPMMC, Universit\'{e} Joseph Fourier and CNRS, 25
Avenue des Martyrs, BP 166, 38042 Grenoble, France}
\author{F.~S.~Bergeret}
\affiliation{Centro de F\'{i}sica de Materiales (CFM-MPC), Centro
Mixto CSIC-UPV/EHU, Manuel de Lardizabal 5, E-20018 San
Sebasti\'{a}n, Spain}
\affiliation{Donostia International Physics Center (DIPC), Manuel
de Lardizabal 4, E-20018 San Sebasti\'{a}n, Spain}
\date{\today}

\begin{abstract}
We investigate heat and charge transport through a  diffusive SIF$_1$F$_2$N tunnel junction, where N (S) is a  normal (superconducting) electrode,
I is an insulator layer and F$_{1,2}$ are   two ferromagnets with arbitrary direction of  magnetization.
The flow of an electric current in such structures at subgap bias is accompanied by a heat
transfer from the normal metal into the superconductor, which enables
refrigeration of electrons in the normal metal.
We demonstrate that the
refrigeration efficiency depends on the strength of the ferromagnetic exchange field $h$ and the
angle $\alpha$ between the magnetizations of the two F layers.   As expected, for  values of $h$ much larger than the superconducting order
parameter  $\Delta$,   the proximity effect is  suppressed and the efficiency of refrigeration increases  with respect to a NIS junction.
However,  for  $h\sim \Delta$  the cooling power (i.e. the heat flow out of the
normal metal reservoir) has  a non-monotonic behavior as a function of $h$ showing a minimum at $h\approx\Delta$.
We also determine the dependence of the cooling power on the lengths of  the ferromagnetic layers, the bias voltage, the temperature, the
transmission of the  tunneling barrier and the magnetization misalignment angle $\alpha$.
\end{abstract}


\maketitle


\section{Introduction}

The presence of the superconducting energy
gap leads to a selective tunneling of high-energy
quasiparticles out of the normal metal in a normal metal - insulator - superconductor (NIS)
tunnel junction.  \cite{Nahum, LPA} This phenomenon
generates a heat current from the normal metal
to the superconductor (also referred to as ``cooling power'').
The heat transfer  through  NIS junctions can be used for the
realization of microcoolers. \cite{Giazotto2006, Review_2, Review_3}
Present state-of-the-art experiments allow the
reduction of the electron temperature in a normal metal lead from 300 to about 100 mK, offering
perspectives for on-chip cooling of nanosized systems, such as high-sensitive
detectors and quantum devices. \cite{Clark, Ullom}

The cooling power  of tunnel junctions depends on several parameters, some of them controllable.
For example the cooling power can be optimized by  controlling the  voltage across the junction. A  maximized cooling effect is reached
at a voltage bias just below the superconducting energy gap $\Delta$.
Larger values of voltage, $eV \gtrsim \Delta$, lead to a larger charge current $I$ through the junction and hence to larger values of  the
Joule heating power, i.e. to a negative  cooling power.
A  limitation of the performance of a NIS microcooler  arises also  from the fact that nonequilibrium
quasiparticles injected into the superconducting electrode accumulate near
the tunnel interface. \cite{Ullom, Pekola_A, VH} As a consequence hot quasiparticles may tunnel  back  into the normal metal,
leading to a reduction of the cooling effect. \cite{VH,Jug}
In order to overcome this problem  a so called quasiparticle trap,\cite{traps} made of an additional
normal metal layer has been attached to  the superconducting electrode, removing hot quasiparticles
from the superconductor.  Recently, it was also shown that a small
magnetic field enhances relaxation processes in a superconductor and leads to significant improvement of the  cooling power
in NIS junctions.\cite{Pekola_H} Improved cooling performance can be also achieved by proper tuning of the
tunneling resistances of the individual NIS tunnel junctions in a double junction SINIS cooling device. \cite{Chaudhuri}

Another important  limitation for NIS microcoolers arises from
the intrinsic multi-particle nature of current transport in NIS junctions which
is governed not only by single-particle tunneling but also by two-particle processes due to the
Andreev reflection.\cite{Andreev} While the single-particle current and the associated heat
current are due to quasiparticles with energies $E > \Delta$,   at low temperatures or high junction transparencies the
charge transport in NIS junctions is dominated by the Andreev reflection, i.e. by subgap processes.
The Andreev current $I_A$ does not transfer heat through the NS interface but rather generates the Joule
heating $I_A V$. At low enough temperatures this heating   exceeds the single-particle cooling.\cite{Sukumar, BA, VB}
The interplay between the single-par\-ticle tunneling and Andreev reflection sets a
limiting temperature for the refrigeration $T_{min}$.\cite{VB}

One way to decrease $T_{min}$ is to decrease the NIS junction transparency.
However, large values of the contact resistance hinder carrier transfer and lead to
a severe limitation in the achievable cooling powers. In order to increase the junction transparency
and at the same time to reduce the Andreev current, it was suggested to use materials where the
proximity effect is suppressed, such as  ferromagnets, ferromagnetic insulators, and half-metals.  In particular  Giazotto \textit{et al.}
studied theoretically a  ballistic normal metal - ferromagnet - superconductor structure within a phenomenological model and predicted
an enhancement  of the cooling efficiency compared to  NIS junctions.\cite{Giazotto}
The  reason for that increase  lies in the suppression of the Andreev reflection due to the band structure of the
ferromagnetic metals. The electron involved in Andreev reflection and its time-reversed
counterpart (hole) must belong to opposite spin bands; thus, suppression of the Andreev
current occurs in a FS junction and its intensity depends on the degree of the electron polarization at the Fermi level which is proportional
to the exchange field of the F layer.\cite{deJong1995,RevG, RevB, RevV}
The enhancement of the cooling efficiency by the magnetic-field-driven tunable
suppression of the Andreev reflection in superconductor/two-dimensional electron gas nanostructures was also studied in Ref.~\onlinecite{Giaprl}.
Note  that   theoretical studies of electron cooling in SF proximity systems were performed only in the ballistic case, \cite{Giazotto, Burmistrova}
while real metallic systems are in the diffusive limit.
Moreover, ferromagnets show in general a  multi-domain structure that was not considered in previous  articles.

In this work we  present a  quantitative  analysis of the thermoelectric transport in NIS microcoolers with a diffusive
ferromagnetic interlayer consisting of two magnetic domains with arbitrary direction of magnetization
(so called ``superconducting triplet spin-valve''\cite{spin-valve}).   Based on the quasiclassical
Keldysh Green functions  formalism  we compute the electric and heat currents through the junction. We show that the enhancement
of the cooling power with respect to the NIS case, as proposed in Ref.~\onlinecite{Giazotto}, only works if the exchange field of the
ferromagnetic interlayer $h$ is much larger than the superconducting order parameter $\Delta$. However, the cooling power shows a
minimum value for $h\approx\Delta$. We also study the dependence of the cooling power on the angle  $\alpha$ between the magnetizations
of the ferromagnetic  domains.  In the case of weak ferromagnets, i.e. for $h\lesssim\Delta$, the antiparallel configuration $\alpha=\pi$ leads to
higher values of the cooling power and smaller values of $T_{min}$. By increasing $h$ this behavior is reversed and the mono-domain configuration ($\alpha=0$)
is more favorable for the refrigeration.  For large values of the exchange field, $h\gg\Delta$, the cooling power is almost independent on $\alpha$.

The paper is organized as follows. In the next section, we formulate the theoretical
model and basic equations.  In particular, we obtain the expressions for the electric and the heat current and identify the
contributions corresponding to single-particle and Andreev tunneling events.
In section III we present and discuss the main results of our work. We finally conclude with a summary of them in section IV.


\section{Model and basic equations}

\begin{figure}[t]
\includegraphics[width=\columnwidth]{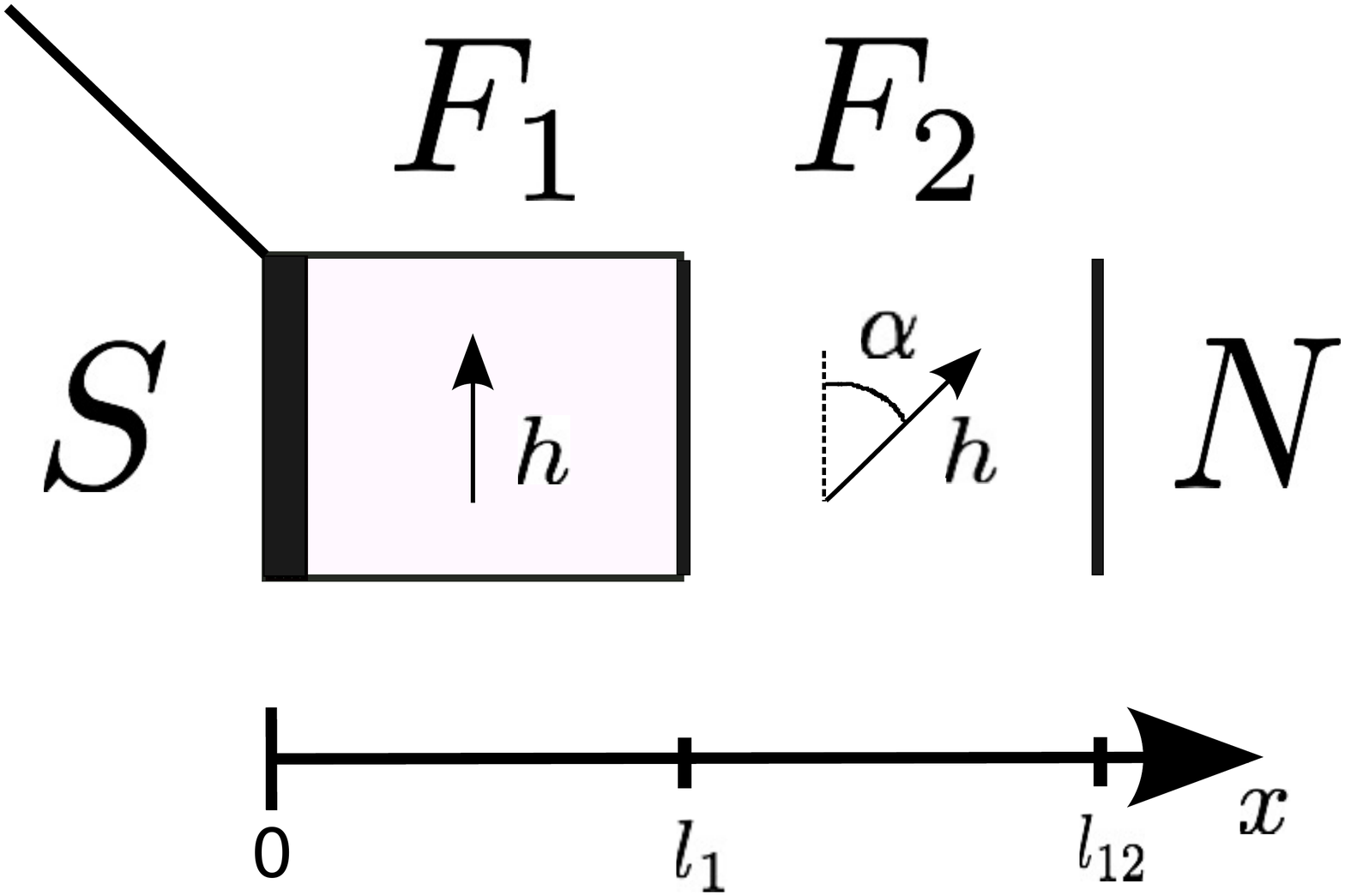}
\caption{The  SIF$_1$F$_2$N junction.
The interface at $x=0$ corresponds to the insulating barrier (thick black line). Interfaces at $x=l_1$ and $x=l_{12}$ are fully transparent.
 $\alpha$ is the relative angle between the magnetization directions of  F$_1$ and F$_2$.} \label{model}
\end{figure}

We consider the SIF$_1$F$_2$N junction depicted in Fig.~\ref{model}.
A ferromagnetic bilayer F$_1$F$_2$ of length $l_{12}=l_1+l_2$ smaller than the inelastic relaxation length\cite{Arutyunov} is connected
to a superconductor (S)  and a normal (N) reservoirs along the $x$ direction.
The  F$_1$F$_2$ bilayer can either model  a two domain ferromagnet or an artificial  hybrid magnetic structure.
We consider the diffusive limit, i.e the elastic scattering length $\ell \ll \min (\xi_h, \xi)$, where $\xi_h = \sqrt{\mathcal{D}/ 2 h}$
is the characteristic penetration length of the superconducting condensate into the ferromagnet, $h$ is the value of the exchange
field, $\xi = \sqrt{\mathcal{D}/ 2 \Delta}$ is the superconducting coherence length and
$\mathcal{D}$ is the diffusion coefficient  (we set $\hbar =k_{B}=1$ and for simplicity we assume the same $\mathcal{D}$ in the whole structure).

We also assume that the F$_1$F$_2$ and F$_2$N interfaces are transparent, while the SF$_1$ is a tunnel interface. Thus, the two ferromagnetic layers
are kept at the same potential as the voltage-biased normal reservoir.
The magnetization of the F$_1$ layer is along the $z$ direction, while the magnetization of the F$_2$ layer forms an angle $\alpha$ with the
one of the layer F$_1$. Both magnetization vectors lie in the $yz$ plane.
Correspondingly, the exchange field vector in the F$_1$ is given by
${\bf h} = (0, 0, h)$, and in the F$_2$ layer
by ${\bf h} = (0, h\sin\alpha, h\cos\alpha)$, where
the angle $\alpha$ takes values from  0 (parallel configuration)
to  $\pi$ (antiparallel configuration).

In order to describe the heat and electric currents through the structure we
introduce the quasiclassical matrix Green function $\breve G$,\cite{LOnoneq,Belzig}
\begin{equation}\label{kelmat}
\breve{G} = \begin{pmatrix} \check{G}^R & \check{G}^K \\
0 & \check{G}^A
\end{pmatrix}.
\end{equation}
The latter is a matrix in the Keldysh $\times$ Nambu $\times$ spin space.
The R, A and K indices stand for the retarded, advanced and Keldysh components (we use the symbols $\breve .$  for $8\times 8$
and $\check .$  for $4\times 4$ matrices). By neglecting non-equilibrium effects,
the Keldysh component is related to the retarded and advanced ones by
\begin{subequations}
\begin{align}
\check{G}^K &= \check{G}^R \check{n} - \check{n} \check{G}^A,
\quad \check{n} = n_+ + \tau_z n_-,
\\
n_\pm &= \frac{1}{2}\left( \tanh \frac{E + eV}{2T_N} \pm
\tanh\frac{E - eV}{2T_N} \right),
\end{align}
\end{subequations}
where $n_\pm$ and $T_N$ are correspondingly the equilibrium quasiparticle distribution functions and the temperature in the normal reservoir
and $\tau_z$ is the Pauli matrix in Nambu space. The retarded and advanced components are related via
$\check{G}^A =-\tau_z \check{G}^{R \dagger} \tau_z$.\cite{LOnoneq}

The matrix \Eq{kelmat} obeys the Usadel equation,\cite{Usadel} which in the notations of Ref.~\onlinecite{Ivanov} reads
\begin{equation}\label{Usadel3D}
i\mathcal{D} \partial_x \breve{J} = \left[ \tau_z\left(E  - {\bf h} {\bf \sigma}\right),
\breve{G}\right], \quad \breve{J} = \breve{G}\partial_x \breve{G}, \quad \breve{G}^2 = 1,
\end{equation}
where ${\bf \sigma} = (\sigma_x, \sigma_y, \sigma_z)$ are the Pauli matrices in spin space.
In the F$_1$ region  ${\bf h \sigma} = h \sigma_z$ and the Usadel equation \Eq{Usadel3D} has the form
\begin{equation}\label{Usadel}
i\mathcal{D} \partial_x \breve{J} = \left[ \tau_z\left(E  - \sigma_z h\right),
\breve{G}\right], \quad \breve{G}^2 = 1.
\end{equation}
In the F$_2$ region ${\bf h \sigma} = h \sigma_z \exp(-i \sigma_x \alpha)$.
It is convenient to introduce Green's functions rotated in spin-space,
\cite{Bergeret2002}
\begin{equation}\label{gauge}
\widetilde{\breve{G}} = U^\dagger \breve{G} U, \quad U =
\exp\left( i \sigma_x \alpha/2 \right).
\end{equation}
The  rotated function $\widetilde{\breve{G}}$ is then determined by \Eq{Usadel}.

The Usadel equation \Eq{Usadel} should be complemented by boundary conditions at the interfaces.
As mentioned above, we assume that
the F$_1$F$_2$ and F$_2$N interfaces are transparent and therefore the boundary conditions at $x=l_1,l_{12}$ read
\begin{eqnarray}
\breve{G} \bigl |_{x=l_1 - 0}& =& \breve{G} \bigl |_{x=l_1 + 0}\label{bc1},\\
\partial_x \breve{G} \bigl |_{x=l_1 - 0}& =& \partial_x \breve{G}
\bigl |_{x=l_1 + 0}\label{bc11},\\
\breve{G}\bigl|_{x=l_{12}-0} &=&\tau_z.
\end{eqnarray}

At $x=0$,  the SF$_1$ interface is a tunnel barrier, and we
may use the Kupriyanov-Lukichev boundary conditions,\cite{KL}
\begin{equation}\label{kupluk}
\breve{J} \bigl |_{x=0} = (W/\xi) \left[
\breve{G}_S, \breve{G} \right]_{x=0},
\end{equation}
where $\breve G_S$ is the Green function of a bulk BCS superconductor defined as
\begin{subequations}
\begin{align}
\breve{G}_S &= \tau_z u + \tau_x v, \label{super}
\\
(u, v) &= (E, \; i\Delta)/\epsilon, \quad \epsilon =
\sqrt{(E+i\eta)^2 - \Delta^2}, \label{uv}
\end{align}
\end{subequations}
where $\eta$ describes inelastic scattering rate and $W\ll 1$ is  the diffusive transparency parameter, \cite{Chalmers}
$W = \xi/ 2 g_N R$. Notice that  $W$  is temperature-dependent, since the coherence length $\xi$
is proportional to  $\Delta^{-1/2}(T)$. In section III  we set $\eta \simeq 10^{-3} \Delta_0$ in our calculations, where $\Delta_0$ is the superconducting gap at $T=0$. In the following we omit $\eta$ in our analytical expressions for simplicity.

Because of the  low transparency of the SF$_1$
barrier, the proximity effect is weak and the  retarded Green function can be linearized (we omit the superscript $R$),
\begin{equation}\label{G_lin}
\check{G} \approx \tau_z + \tau_x \hat{f},
\end{equation}
where $\hat{f}$ is the $2 \times 2$ anomalous Green function in the spin space ($|\hat{f}|\ll 1$) that obeys  the linearized Usadel equation,
\begin{equation}\label{linearized}
i\mathcal{D} \partial^2_{xx} \hat{f} = 2 E \hat{f} - \left\{ {\bf h
\sigma}, \hat{f} \right \},
\end{equation}
where $\{\cdot, \cdot\}$ stands for the anticommutator.

The general solution of \Eq{linearized} has the form
\begin{equation}\label{f}
\hat{f}(x) = f(x) + f_y(x) \sigma_y + f_z(x) \sigma_z,
\end{equation}
where $f$ is the singlet component and $f_z$, $f_y$ are the triplet components with respectively zero and $\pm 1$ projections on the spin
quantization axis (we choose the $z$-axis).
The $f_y$ term is usually known as the long range triplet component because
it describes Cooper pairs with parallel spins  which survive the strong exchange splitting and can diffuse
into the ferromagnet over larger distances compared to the singlet component.\cite{Bergeret, Klapwijk, Blamire, Birge, Robinson}
Indeed, by substituting   Eq.~(\ref{f}) into Eq.~(\ref{linearized}) and using the boundary conditions Eqs.~(\ref{bc1}-\ref{kupluk}) one can
compute the components $f(x)$, $f_y(x)$ and $f_z(x)$ ,
and easily show that while $f(x)$ and $f_z(x)$ decay into the ferromagnet over the  magnetic length $\sqrt{\mathcal{D}/2h}$
the long-ranged component $f_y(x) $ decays over the  length given by $\sqrt{\mathcal{D}/2E}$.

The charge  and energy  currents,  $I$ and $Q$ respectively,  can be obtained from\cite{LOnoneq, Belzig, Vinokur, Golubov}
\begin{subequations}
\begin{align}\label{I}
I &= \frac{g_N}{e} \int_0^{\infty} I_- \; dE, \quad Q =
\frac{g_N}{e^2} \int_0^{\infty} E I_+ \; dE,
\\
\label{Ipm} I_- &\equiv (1/8) \tr \tau_z \check{J}^K, \quad I_+
\equiv (1/8) \tr \tau_0 \check{J}^K,
\\
\check{J}^K &\equiv \left( \breve{G}\partial_x \breve{G} \right)^K
= \check{G}^R \partial_x \check{G}^K + \check{G}^K \partial_x \check{G}^A.
\end{align}
\end{subequations}
where  $\breve{J}^K$ is the Keldysh component of the matrix current  defined in Eq. (\ref{Usadel3D})
and $\tau_0$ is the unitary matrix in Nambu space.

From  Eqs. (\ref{kupluk}), (\ref{super}) and (\ref{f}) we determine both currents, $I$ and $Q$ at the SF$_1$ interface
\begin{subequations}\label{IQ}
\begin{align}
I &= \frac{1}{eR} \int_0^\infty n_- \left( \re u + \re v \re f_0 \right) \; dE,\label{II}
\\
Q &= \frac{1}{e^2 R} \int_0^\infty E (n_+ - n) \left( \re u - \im v \im f_0 \right) \; dE,\label{QQ}
\end{align}
\end{subequations}
where $n = \tanh (E/ 2T_S)$ and $T_S$ are  the equilibrium quasiparticle distribution function and temperature in the superconducting
reservoir respectively.   The function $f_0 \equiv f \bigl |_{x = 0}$ is the singlet component of $\hat{f}$ at $x = 0$.
Notice that only the singlet component of $\hat{f}$ enters the equations for the electric and energy currents.
There is however an indirect dependence of the currents on the triplet component  since the amplitude of $f_0$ in turn  depends on the amplitudes
of the triplet $f_y$ and $f_z$.

We now examine \Eqs{IQ} and discuss separately two main contributions for the currents $I$ and $Q$, which
originate from the single-particle ($E > \Delta$) and Andreev ($0 < E < \Delta$) processes.
Let us focus first on the single particle contributions
$I_1$ and  $Q_1$.
For energies larger than the superconducting gap ($E > \Delta$) only the terms proportional to $ \re u$ in \Eqs{IQ} are non-zero.
From  \Eq{uv} and \Eq{II} we obtain  the  single-particle contribution to the  electric current
\begin{equation}\label{I1}
I_1 = \frac{1}{eR} \int_\Delta^\infty N_S(E) n_-(E) \; dE,
\end{equation}
where $N_S(E) = |E| \Theta(|E| - \Delta)/\sqrt{E^2 - \Delta^2}$ is the BCS normalized density of states (DOS) and
$\Theta(x)$ is the Heaviside step function.

Rewriting in \Eq{I1} the $n_-(E)$ function in terms of the Fermi function in the N reservoir,
$n_F(E)=[1+\exp(E/T_N)]^{-1}$,
we arrive at the well known expression for the tunneling current, \cite{Werthamer}
\begin{align}
I_1 = \frac{1}{eR}\int_{-\infty}^{\infty} N_S(E) [
n_F(E-eV)-n_F(E)]\,dE.\label{IT}
\end{align}
Note that  within the linear approach the normalized DOS in the F layer is equal to unity
and therefore the single particle electric current is  independent of $f_0$.

The single-particle contribution to the energy current can be obtained from \Eq{QQ},
\begin{equation}\label{Q1}
Q_1 = \frac{1}{e^2 R} \int_\Delta^\infty E(n_+ - n) \left[ N_S(E) - M_S^+(E) \im f_0 \right] \; dE,
\end{equation}
where $M_S^+(E) = \Delta \Theta(|E| - \Delta)/\sqrt{E^2 - \Delta^2}$.

For energies $E < \Delta$ the electric charge is transferred by means of the Andreev reflection.
The subgap current or Andreev current can be  obtained from \Eq{II},
\begin{equation}\label{IA}
I_A = \frac{1}{eR} \int_0^\Delta n_-(E) M_S^-(E) \re f_0 \; dE.
\end{equation}
where $M_S^-(E) = \Delta \Theta(\Delta - |E|)/\sqrt{\Delta^2 - E^2}$.\cite{footnote1}

According to \Eq{QQ} the contribution of the Andreev processes to the  energy current vanishes, $Q_A = 0$.

We are interested here in the cooling power $P$, i.e. in
the heat current flowing out of the normal metal reservoir.
One can express the cooling power in terms of the contributions introduced previously, \cite{VB}
\begin{equation}\label{P}
P = - Q - I V = P_1 + P_A,
\end{equation}
where
\begin{equation}
P_1 = - Q_1 - I_1 V, \quad P_A = - I_A V.
\end{equation}

From Eqs.~(\ref{IT},\ref{Q1},\ref{IA}) it is clear that for equal temperature of the electrodes ($T_N = T_S$) and no
bias voltage the cooling power vanishes.  For a finite voltage $V$,  on one hand the heat is taken from the N
reservoir and is released in the superconductor. On the other hand  there is a global  heat production in both electrodes
due to the Joule heating.  In particular the  Andreev current $I_A$ contributes to the  Joule heating $I_A V$, which is
fully released in the normal metal electrode and leads  to a reduction of the cooling power.

In the next section we calculate the cooling power of the SIF$_1$F$_2$N junction
as a function of  the different parameters by solving  Eqs.~(\ref{IT},\ref{Q1},\ref{IA},\ref{P}).


\section{Results and discussion}

According to  \Eqs{IQ}, in the linear case both the electric and heat currents are determined by
the singlet component  $f_0$ of the anomalous Green function,
\Eq{f}, evaluated at the SF$_1$ interface ($x=0$).
Solving \Eq{linearized} in the F$_1$ layer we obtain for the components of \Eq{f},
\begin{subequations}\label{f_i}
\begin{align}
f_\pm(x) &= a_\pm \cosh (k_\pm x) + \frac{2W}{k_\pm} (u a_\pm - v)
\sinh (k_\pm x),
\\
f_y(x) &= a_y \cosh (k_y x) + \frac{2W}{k_y} u a_y \sinh (k_y x),
\end{align}
\end{subequations}
where $f_\pm = f \pm f_z$,  $a_i$ are the boundary values of $f_i$ at  $x=0$
($i$ stands for $+,-,y$) and
\begin{equation}
k_\pm = \sqrt{\frac{2(E \mp h)}{i\mathcal{D}}}, \quad
k_y =\sqrt{\frac{2E}{i\mathcal{D}}}.
\end{equation}
In the F$_2$ layer  the general solution has the form,
\begin{equation}
\widetilde{f}_i(x) = b_i \sinh \left[ k_i (x - l_{12}) \right],
\end{equation}
where $\widetilde{f}_i$ are the components of the rotated Green function, \Eq{gauge}.

Using the boundary conditions at the   F$_1$F$_2$ interface, Eqs. (\ref{bc1}-\ref{bc11}) one obtain a set of six linear equations for the
six coefficients $a_i$ and $b_i$, that can be solved straightforwardly.

\begin{figure}[tb]
\includegraphics[width=\columnwidth]{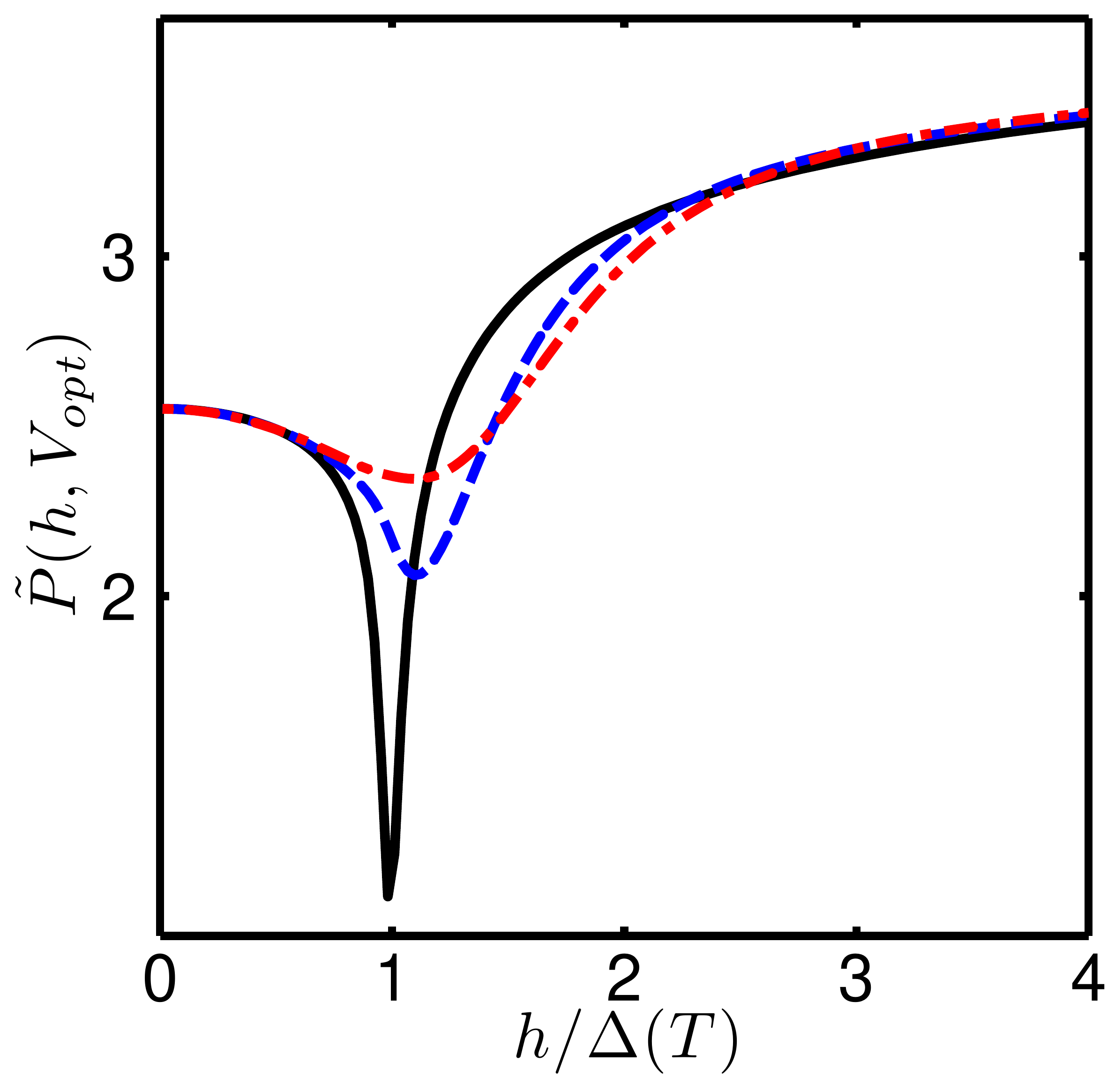}
\caption{(Color online) Cooling power versus exchange field for different
orientations of the exchange field vector in the second ferromagnetic layer F$_2$: $ \alpha= 0$ (black solid line),
$ \alpha= \pi /2$ (blue dashed line) and $ \alpha= \pi $ (red dash-dotted line), calculated at optimum bias;  $W=7 \times 10^{-3}$,
$T=0.25 \Delta_0$, $l_1=\xi$ and $l_2=6 \xi$.  We have defined $\tilde P=10^{2}P(V_{opt})e^2R_0/\Delta^2_0$.} \label{Pvsh10}
\end{figure}

In particular we are interested in the value of the singlet component of the anomalous Green function
at $x=0$ which is given by  $f_0=(a_++a_-)/2$.
The analytical expression has lengthy awkward form and we do not present it here.
Once we obtain $f_0$ we compute the charge and energy currents from  \Eqs{I1}, \eqref{Q1}
and \eqref{IA}. Finally, using \Eq{P} we determine the cooling power.
In what follows we assume that the temperatures of the S and
N reservoirs to be equal, $T_S = T_N = T$, and neglect nonequilibrium effects in the ferromagnetic interlayer.\cite{VB}

\begin{figure}[t]
\includegraphics[width=\columnwidth]{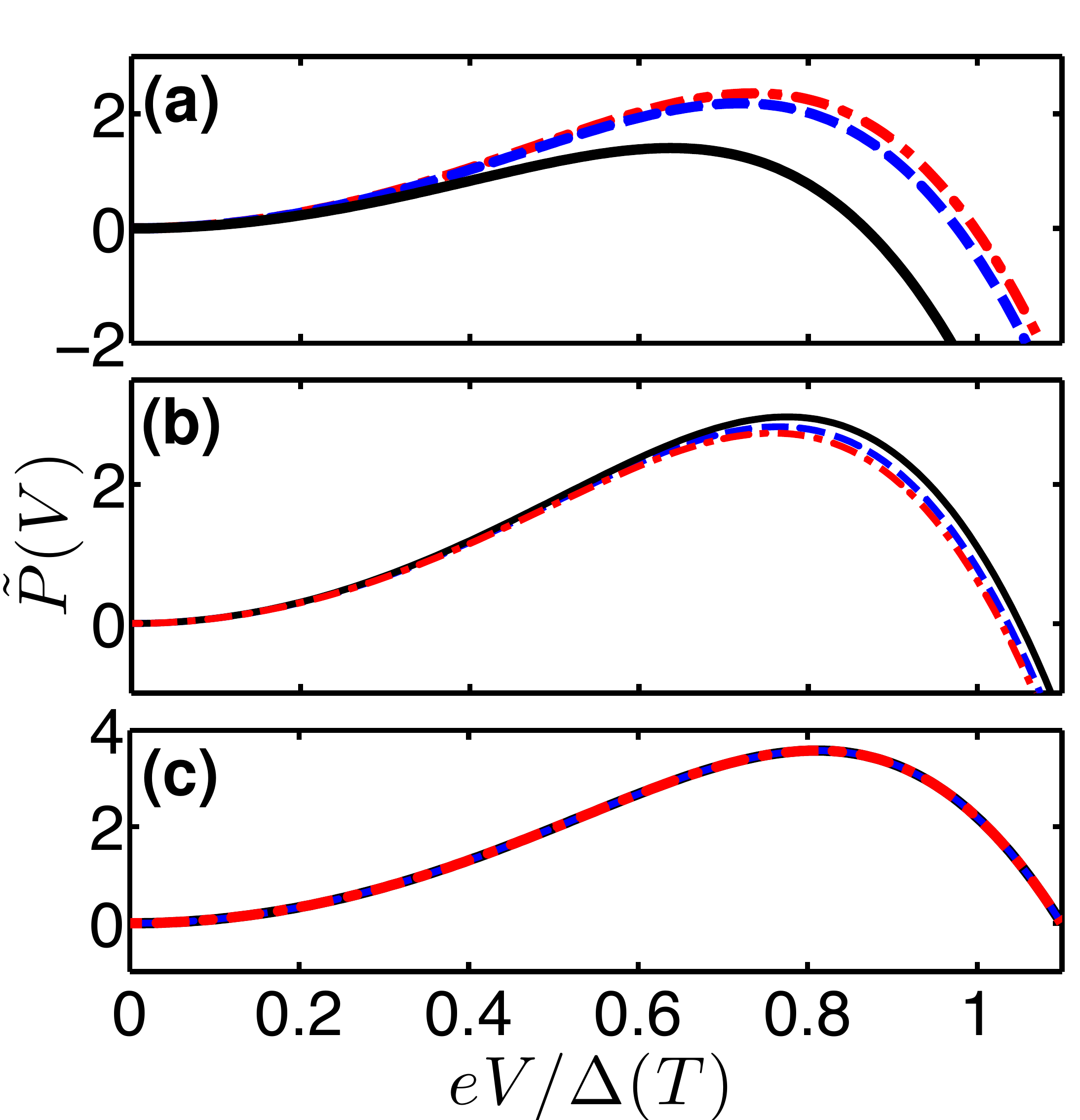}

\caption{(Color online) Cooling power versus bias voltage  for $h=\Delta(T)$ (a),
$h=1.7\Delta(T)$ (b) and $h=8\Delta(T)$ (c)  for different
orientations of the exchange field vector in the second ferromagnetic layer F$_2$: $\alpha= 0$ (black solid line),
$\alpha= \pi /2$ (blue dashed line) and $\alpha= \pi$ (red dash-dotted line);
$W=7 \times 10^{-3}$, $T=0.25 \Delta_0$, $l_1=\xi$ and $l_2=6 \xi$. $\tilde P$ is defined in Fig. 2.} \label{panelv}
\end{figure}

The bias voltage between the S and N reservoirs is an easily adjustable
experimental parameter, so all our curves except those presented in Fig.~\ref{panelv} are calculated for optimal value
of the voltage bias $V_{opt}$, at which the cooling power reaches its maximum for given values of the
other parameters. In what follows, we assume the quantity $W$ to be taken at $T=0$, allowing for its temperature
dependence in \Eqs{f_i} by means of corresponding temperature-dependent factors.
In the subsequent analysis the cooling power $P$ is given in units of  $\Delta^2_0/e^2 R_0$, where
$\Delta_0$ is the value of $\Delta$ at zero temperature and $R_0$ is the junction resistance at a fixed value $W=10^{-2}$ of the
tunneling parameter.

\begin{figure}[t]
\includegraphics[width=\columnwidth]{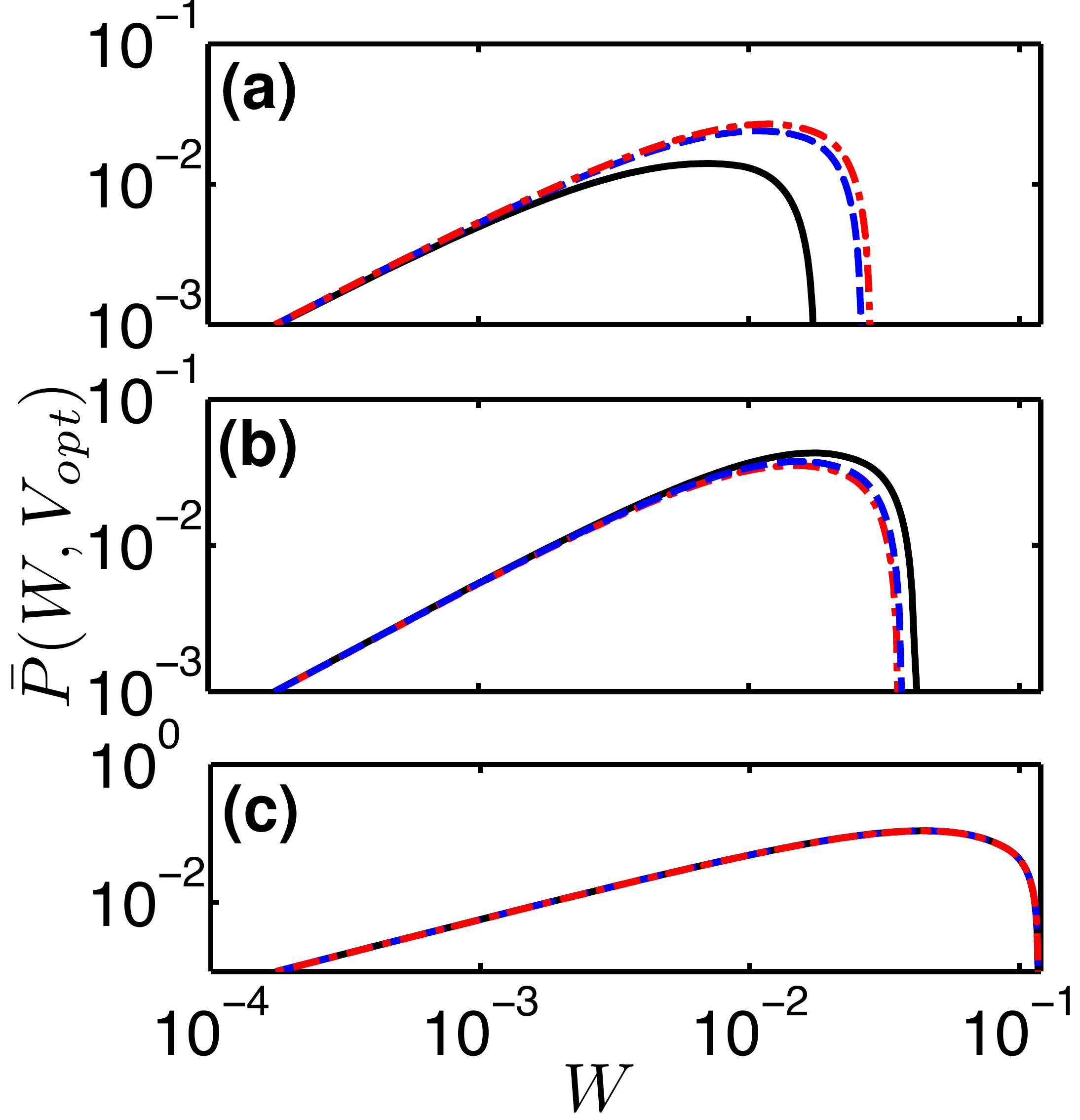}

\caption{(Color online) Dependence of the cooling power on the tunneling parameter $W$ for
$h=\Delta(T)$ (a), $h=1.7\Delta(T)$ (b) and $h=8\Delta(T)$ (c) and for different
orientations of the exchange field vector in the second ferromagnetic layer F$_2$: $\alpha= 0$ (black solid line),
$\alpha= \pi /2$ (blue dashed line) and $\alpha= \pi$ (red dash-dotted line). $P$ is calculated at optimum bias;
$T=0.25 \Delta_0$, $l_1=\xi$ and $l_2=6 \xi$. We have defined $\bar P(W,V_{opt})=P(W,V_{opt})e^2R_0/\Delta^2_0$.
Note the logarithmic scale.} \label{panelW}
\end{figure}

We first  study the dependence of the cooling  power on the strength of the exchange field $h$.
This dependence is shown in  Fig.~\ref{Pvsh10} for three different angles $\alpha=0,\pi/2,\pi$ between
the magnetizations of F$_1$ and F$_2$ layers at the optimum value of bias voltage. We have chosen the
values of the temperature and tunneling parameter $W$ such that the Andreev current role in the cooling
processes is essential (see Fig.~\ref{panelW}).\cite{VB} The thickness of the F layers is chosen to be $l_1=\xi $ and $l_2=6\xi$.

Depending on the value of $l_1/\xi_h$, where  $\xi_h=\sqrt{\mathcal{D}/2h}$ is the characteristic  penetration length of the
superconducting condensate into F$_1$, one identifies different behaviors.  If $l_1\gg\xi_h$, i.e. for large values of $h/\Delta(T)$
the amplitude of the superconducting condensate  in F$_2$ can be neglected, as well as  the dependence of the $f_0$ function on the angle $\alpha$.
Thus, in the limit $h/\Delta(T)\gg1$,   the  value of the cooling power  does not depend on $\alpha$. Moreover, this asymptotic value  is larger than
in the nonmagnetic case ($h=0$).  This is a consequence of the strong suppression  of the singlet correlations in F$_1$ due to the exchange field and
hence of the Joule heating associated to the  Andreev current [see \Eq{P}].  Note that for the value of temperature used in
our figures $\Delta(T)\approx\Delta_0$ .

\begin{figure}[tb]
\includegraphics[width=\columnwidth]{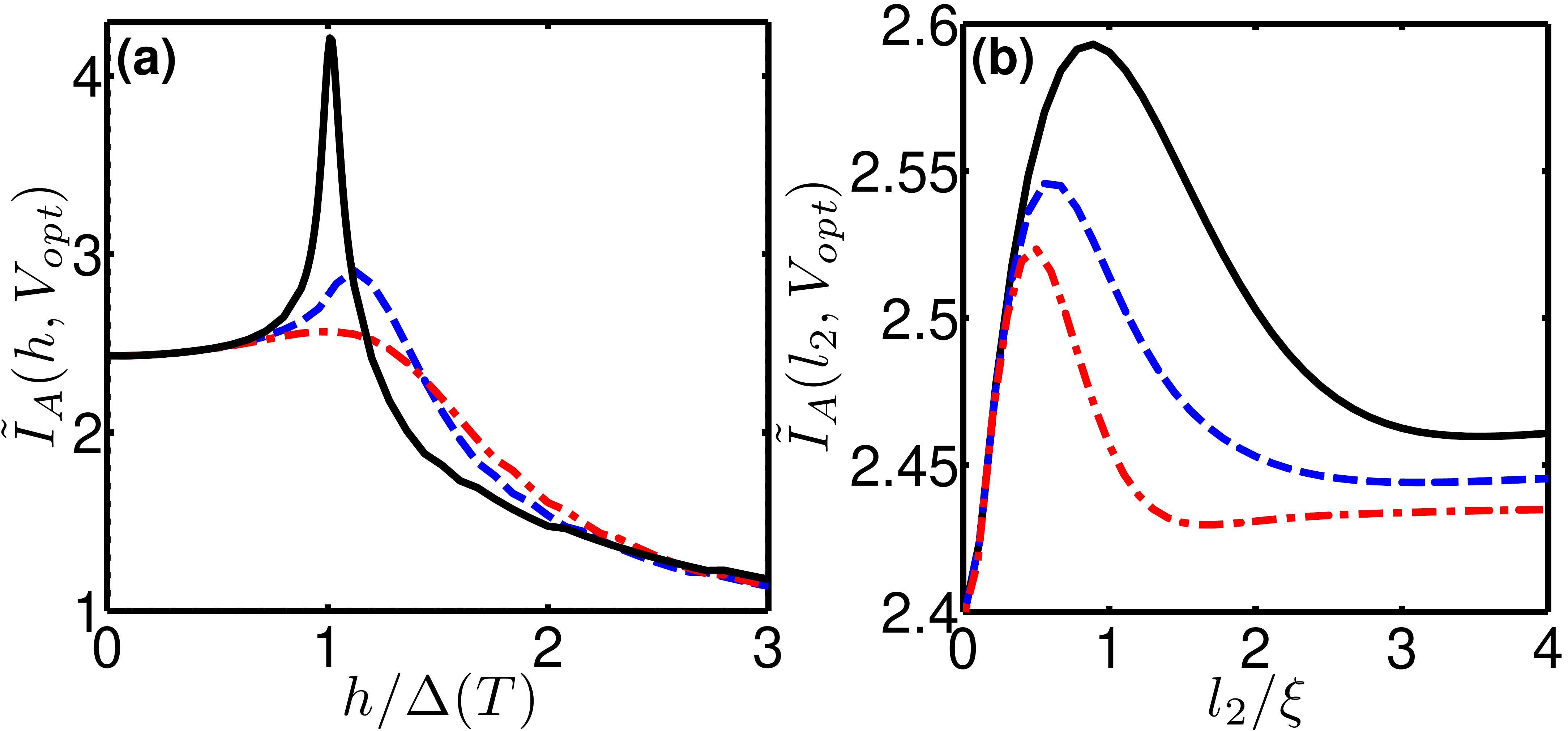}
\caption{(Color online) The Andreev current as a function of (a) the exchange field for  $l_2 = 6\xi$ and  as a function of (b)  the F$_2$  length  for
$h = 0.7 \Delta(T)$.  Different magnetic configurations are chosen:
$\alpha= 0$ (solid black line), $\alpha= \pi /2$ (dashed blue line),  $ \alpha= \pi $ (dash-dotted red line). The Andreev current is calculated at optimal bias;
$W = 7 \times 10^{-3}$, $T=0.25 \Delta_0$, $l_1=\xi$.
We have defined $\tilde I_A=I_A(V_{opt})eR_0/\Delta_0$.} \label{panelIA}  \vspace{-4mm}
\end{figure}

In the opposite limit, $l_1/\xi_h\ll 1$, the characteristic penetration length  depends weakly on $h$, and therefore the cooling power is also
$\alpha$-independent. However, by increasing $h$   the cooling power first decreases and reaches  a minimum.
This unexpected behavior is qualitatively similar for all  magnetic configurations and is a consequence of the Andreev current peak at $h \approx \Delta(T)$
(for mono-domain case) in the  finite temperature and finite voltage regime, see Fig.~\ref{panelIA}(a), solid black line.
However,  there are quantitative differences between the mono-domain  ($\alpha=0$) and two domain  ($\alpha=\pi,\pi/2$) configurations.
For  $\alpha=0$,   $P(h)$ shows a minimum at $h \approx \Delta(T)$.   It is worth mentioning that   around
this minimum the cooling power of the SIF$_1$F$_2$N system is  lower than that of the NIS junction ($h=0$).
By increasing the angle $\alpha$ the minimum is less pronounced and shifts to larger values of $h\gtrsim\Delta(T)$. For these values of $h$
and for $l_1=\xi$ the superconducting condensate can penetrate both ferromagnetic layers.
Thus, the effective exchange field  $\bar h$ acting on the  Cooper pairs is a field, averaged over the length $\xi_h$.\cite{Bergeret_h}
The $\bar h(\alpha)$ is gradually reduced as $\alpha$ increase from $0$ to $\pi$. As before the
cooling power minimum  is at $\bar h (\alpha) \approx \Delta(T)$ which in the case of a finite $\alpha$ corresponds to larger values of the bare $h$.  The minimum
of the cooling power (Fig.~\ref{Pvsh10}), corresponds to   a  maximum of the Andreev current  [Fig.~\ref{panelIA}(a)].  The  unexpected nonmonotonic  behavior of the Andreev current at small exchange fields $h \sim \Delta(T)$ is due to the competition between two-particle tunneling processes and decoherence mechanisms as quantitatively explained in a recent work by the authors. \cite{OVHB}

\begin{figure}[tb]
\includegraphics[width=\columnwidth]{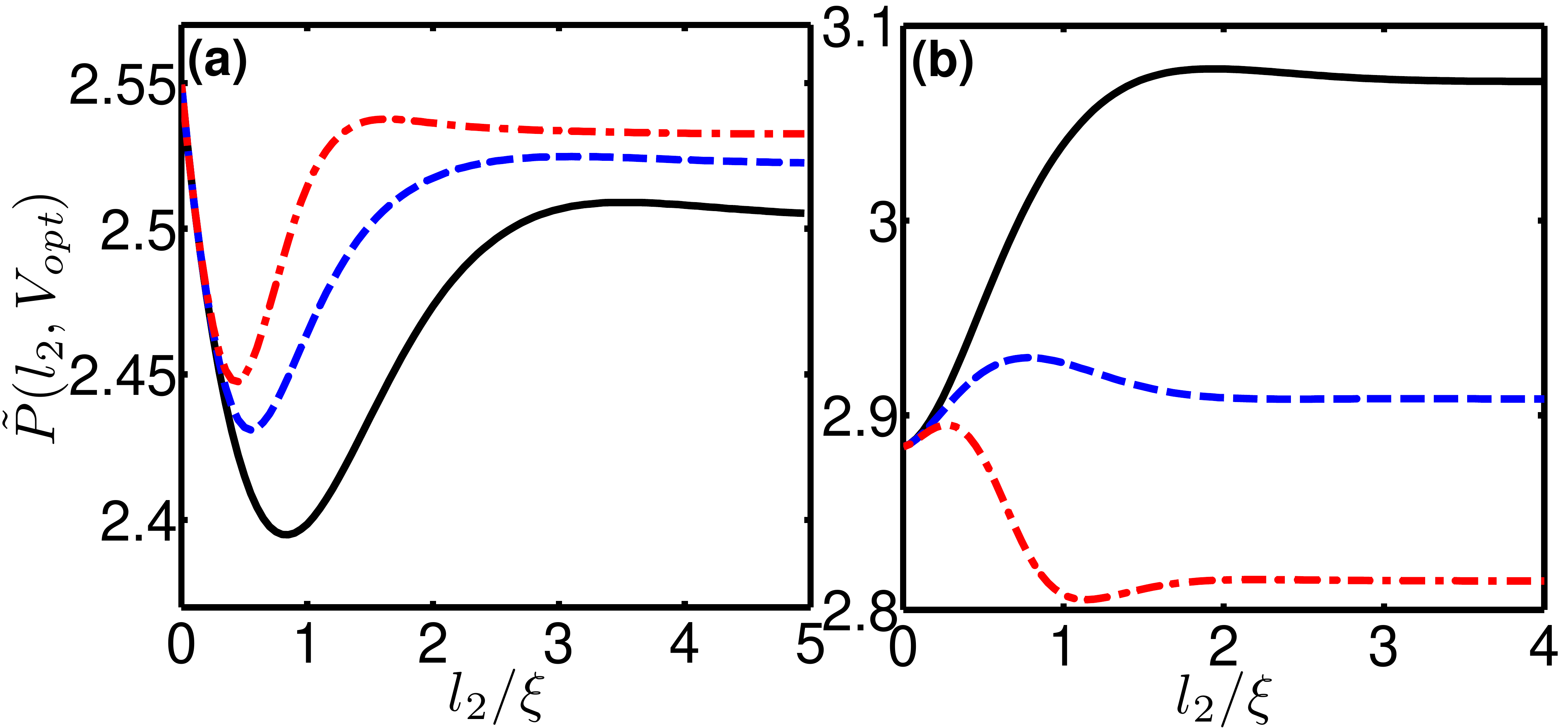}
\caption{(Color online) Cooling power versus  length $l_2$ of  the F$_2$ layer  for (a)  $h=0.7\Delta(T)$ and (b) $h=1.7\Delta(T)$.
We  consider different orientations of the exchange field vector in the second ferromagnetic layer F$_2$ with respect to the one in F$_1$:
$\alpha= 0$ (solid black line), $\alpha= \pi /2$ (dashed blue line),  $ \alpha= \pi $ (dash-dotted red line), and
calculate the cooling power  at optimal bias;
$W = 7 \times 10^{-3}$, $T=0.25 \Delta_0$ and $l_1=\xi$. $\tilde P$ is defined in Fig. 2.} \label{panell2}  \vspace{-4mm}
\end{figure}

We analyze now the dependence of the cooling power on the length of the ferromagnetic bridge F$_1$F$_2$. To do this, we fix the thickness of F$_1$ at
$l_1=\xi$ and vary  $l_2$. Fig.~\ref{panell2} shows the $P(l_2)$ dependence for  two different values of the exchange field $h/\Delta(T)=0.7, 1.7$ and
different magnetic configurations $\alpha=0,\pi/2,\pi$. As expected   all curves tend to a finite asymptotic value  when $ l_2 \gg \xi$.
This value however depends on $\alpha$.

In the case of an exchange field smaller than the superconducting gap  [$h=0.7\Delta(T)$, see Fig.~\ref{panell2}(a)]
the cooling power first reduces monotonically to a minimum by increasing $l_2$, then enhances to a maximum and finally reduces to the asymptotic value.
Such behavior is preserved for all magnetic configurations and it follows from the nonmonotonic  behavior of the Andreev current, shown in Fig.~\ref{panelIA}(b).
Decrease of the Andreev current corresponds to the increase of the cooling power and vice versa. As shown in Fig.~\ref{panelIA}(b),  at large values of $l_2$ the Andreev current  increases
by decreasing $l_2$, reaches a maximum and finally decreases for $l_2 \lesssim \xi$. The strong suppression of the Andreev current for small values of $l_2$  is due to the proximity of the N reservoir at $x = l_{12}$.
On the other hand for larger values of $l_2$ the superconducting proximity effect in the ferromagnetic bridge is fully developed  and leads to an increase of the Andreev current. It is remarkable that the cooling power for $\alpha=\pi$ is larger than the one at $\alpha=0$ for all values of $l_2$.
In this case a lower effective exchange field $\bar h$ leads to larger values of the cooling power, due to the shift of the
minimum of  $P(h)$  observed in Fig.~\ref{Pvsh10}.

For an exchange field larger than $\Delta(T)$ [$h=1.7\Delta(T)$, see Fig.~\ref{panell2}(b)] the behavior
of the cooling power as a function of $l_2$ strongly depends on $\alpha$. For a mono-domain magnet, $\alpha=0$,
the cooling power increases monotonically by increasing $l_2$ until it reaches the asymptotic value due to the suppression
of the Andreev current as in the ballistic case studied in Ref.~\onlinecite{Giazotto}.  Similarly, in the antiparallel
configuration ($\alpha=\pi$), the cooling power first  increases by increasing $l_2$, however for a larger value of $l_2$
reaches a maximum and then decreases.  The presence of F$_2$ with a magnetization antiparallel to the one of F$_1$ leads
to a reduced effective exchange field of the F$_1$F$_2$ bridge.  Thus, the Andreev current contribution is enhanced with respect
to the one in the case $l_2=0$.  As intuitively expected the cooling power (Andreev current) reaches a minimum (maximum)
when $l_2\sim l_1=\xi$, i.e. when the average magnetization is minimized. Further increase of $l_2>\xi$ leads to a
suppression of the Andreev current and therefore to an increase of $P$ until the asymptotic values are reached.
Fig.~\ref{panell2}(b) also shows the intermediate case $\alpha=\pi/2$.

\begin{figure}[tb]
\includegraphics[width=\columnwidth]{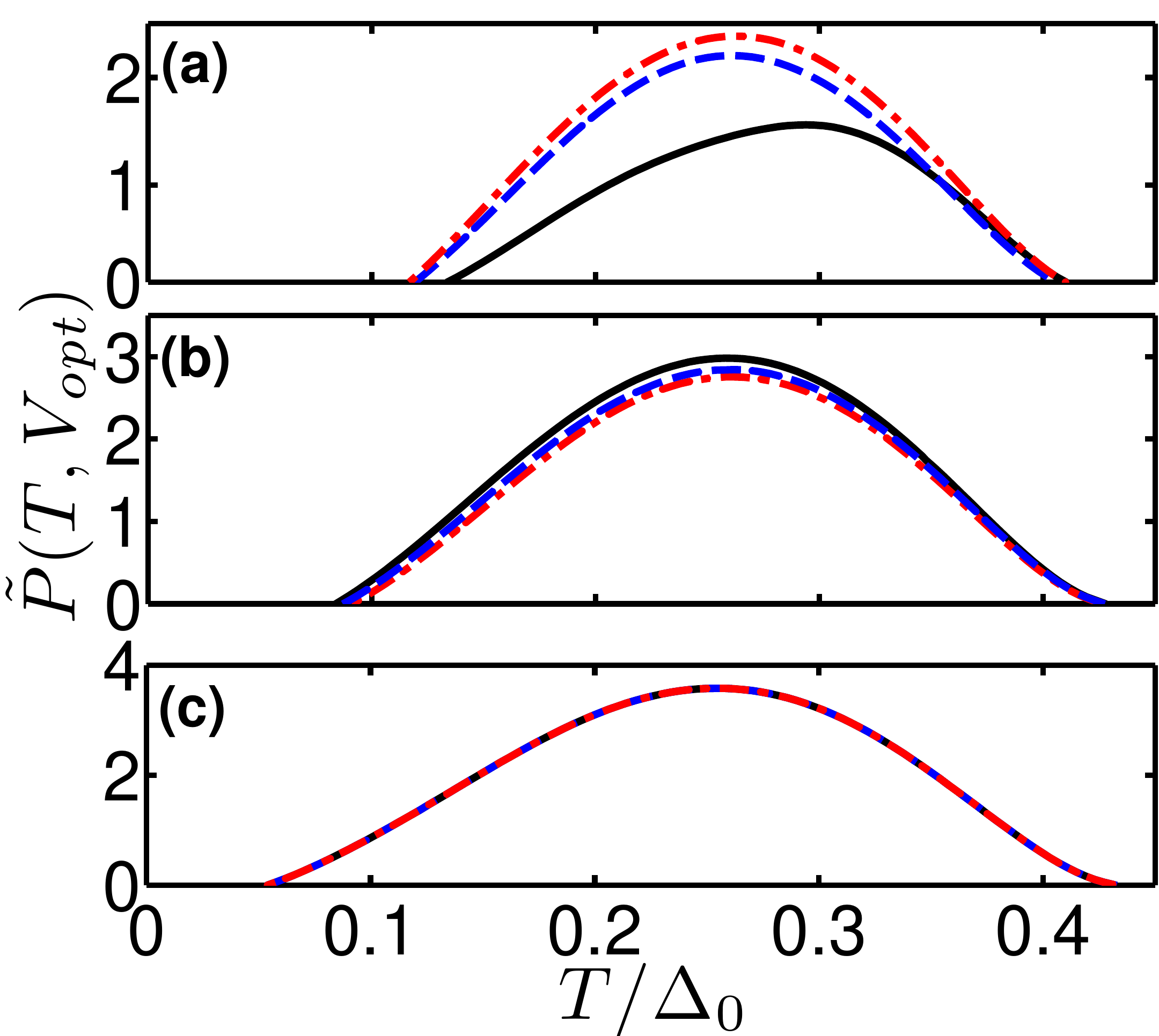}
\caption{(Color online) Temperature dependence of the cooling power for $h=\Delta_0$ (a), $h=1.7\Delta_0$ (b) and $h=8\Delta_0$ (c) and for different
orientations of the exchange field vector in the second ferromagnetic layer F$_2$:
$\alpha= 0$ (black solid line),
$\alpha= \pi /2$ (blue dashed line) and $\alpha= \pi$ (red dash-dotted line). $P$ is  calculated at optimum bias;
$W=7 \times 10^{-3}$, $l_1=\xi$ and $l_2=6 \xi$. $\tilde P$ is defined in Fig. 2.} \label{panelT}
\end{figure}

We now analyze the dependence of the cooling power on the bias voltage $eV$, tunneling parameter $W$ and temperature $T$.
In our subsequent analysis we consider  three different values of the exchange field
$h=\Delta(T),1.7\Delta(T), 8\Delta(T)$, and  three magnetic configurations $\alpha=0,\pi/2,\pi$.
We set $l_1=\xi$, short enough for the pair correlation
to be substantial in the F$_2$ layer [for $h=\Delta(T),1.7\Delta(T)$] and $l_2=6\xi$,  long enough to ensure
the asymptotic regime [see Fig.~\ref{panell2}].
Figs.~\ref{panelv}, \ref{panelW} and  \ref{panelT}  show  the cooling power as a function of
$eV$, $W$ and $T$.    A common  feature of these figures is that the   range  of values of  $V$, $W$ and $T$,  for which the cooling
power is positive increases by increasing $h$.
Also the magnitude of the cooling power increases  with $h$.  This is in agreement with the qualitative predictions of Ref.~\onlinecite{Giazotto}.
Note that the  shape of all  curves in  Figs.~\ref{panelv}, \ref{panelW} and  \ref{panelT}
does not depend significantly on  the angle $\alpha$.

Figs.~\ref{panelv} and \ref{panelW} show that for low values of $eV$ and $W$, respectively, the cooling power depends only
weakly  on the relative magnetization  angle $\alpha$.  However, by increasing $eV$ and $W$ the difference becomes appreciable,
in particular for $h\approx\Delta(T)$.

As shown in Fig.~\ref{panelv} at certain value of $eV_{opt}\lesssim 0.8\Delta(T)$,   the cooling power reaches its
maximum value $P_{max} = P(V_{opt})$. The $eV_{opt}$ value  is the one used as optimal bias value in the figures.
For voltages larger than this optimal value, the quasiparticle current $I$ and hence the Joule heating power
$IV$ increase drastically  leading to a rapid decrease of the cooling power. As can be seen from Figs.~\ref{panelv}, \ref{panelW} and  \ref{panelT}
the optimal voltage $V_{opt}$ depends on the temperature $T$, tunneling parameter $W$ and magnetic configuration angle $\alpha$.
For the exchange field equal to the superconducting gap the maximal cooling power $P_{max}$ is largest in the antiparallel
configuration, while for larger $h = 1.7 \Delta(T)$ the largest value $P_{max}$ is in the parallel configuration, in agreement with Fig.~\ref{Pvsh10}.

Fig.~\ref{panelW} shows that the cooling power has also a maximum as a function of $W$. Increasing $W$ the cooling power first linearly increases as single electron tunneling dominates.
For larger values of the tunneling parameter, the Andreev current heating dominates over the single-particle cooling and leads to a rapid decrease of the
cooling power, which tends to zero at a certain onset point. As the exchange field increases, the role of Andreev processes becomes less important, therefore
the onset shifts towards larger values of $W$. This means that for higher exchange field in the ferromagnetic interlayer one may use weaker tunnel barriers
for the microcooler fabrication, which leads to higher amplitudes of the cooling power [see Fig.~\ref{panelW} (c)] and more effective electron refrigeration.

In Fig.~\ref{panelT} we show the temperature dependence of the cooling power. At $T \gtrsim 0.42 \Delta_0 \approx 0.75  T_c$, where $T_c$ is the critical temperature of the superconductor,
the cooling power
becomes negative for all voltages. This value of the temperature holds for  a wide range of parameters.\cite{VB} The
existence of  such a maximal temperature is due to the increase of the number
of thermally excited quasiparticles which produce
enhanced  Joule heat. By lowering the  temperature  the cooling power at optimal bias
first increases and reaches a maximum.  At lower temperatures, the Joule heat due to
Andreev processes causes the cooling power to decrease. At a certain
temperature $T_{\textit{min}}$, the cooling power tends to zero, which defines
the lower limiting temperature for the cooling regime. As follows from
Fig.~\ref{panelT}, the temperature $T_{\textit{min}}$ decreases when increasing the
exchange field; this is because the Andreev current and the associated Joule heat
are suppressed by the exchange interaction in the ferromagnet.
Finally, one can see from Fig.~\ref{panelT}  that the minimum cooling temperature in the parallel $ T_{min}^P$ and antiparallel
$T_{min}^{AP}$  configuration satisfy:   $T_{min}^{AP} < T_{min}^P$ for $h=\Delta(T)$, while $T_{min}^{AP} > T_{min}^P$ for $h=1.7\Delta(T)$.
For $h=8\Delta(T)$ [Fig.~\ref{panelT}(c)]  $P(T)$ is almost independent on $\alpha$.

A common feature of Fig.~\ref{panelv}, \ref{panelW} and \ref{panelT} is that for rather small value of the exchange field,  $h=\Delta(T)$,  the antiparallel
configuration is more favorable for  cooling [see (a) panels]. For larger exchange field $h = 1.7\Delta(T)$, on the contrary, the
parallel configuration is favorable for cooling [see (b) panels].
As expected, in the case of strong enough ferromagnet [$h=8\Delta(T)$]  the thickness of F$_1$ layer $l_1 \gg \xi_h$ and the superconducting condensate practically does not penetrate into F$_2$ layer. Thus the cooling power is  $\alpha$-independent [see (c) panels of Figs.~\ref{panelv}, \ref{panelW} and \ref{panelT}].

\section{Conclusions}

We have developed a quantitative theory of charge and heat transport in normal metal - superconductor
tunnel junctions with an intermediate ferromagnetic bilayer. We have assumed that the magnetizations
of the ferromagnets form an angle $\alpha$ and  focused our study on  the cooling power of such a structure.

In the previous works it has been suggested that  the larger the exchange field the more efficient the
cooling.\cite{Giazotto}  In this case the enhancement of the cooling is due to the suppression of the
Andreev processes and therefore suppression of the Joule heating, released in the normal metal electrode.
However, our results have shown that this hypothesis is only valid  in the case of strong ferromagnets
[$h\gg\Delta$].  For weak ferromagnets with an exchange field  comparable to the superconducting
order parameter $\Delta$  the cooling power shows a non-monotonic dependence on $h$, with a minimum
at $h\approx\Delta$ (in mono-domain case) that corresponds to a maximum in the Andreev current $I_A$. Moreover, around this
minimum the cooling power of the SIF$_1$F$_2$N structure is even lower than the one of the NIS junction.
We have also shown that in the two-domain case, a finite value of  $\alpha$ shifts  the minimum  of cooling power to larger values of $h$ if the thickness of F$_1$ is comparable to the magnetic length $\xi_h$.  In this case,  the effective exchange field $\bar h$ acting
on the Cooper pairs is gradually reduced as $\alpha$ increases from $0$ to $\pi$. The minimum then is
at $\bar h \approx \Delta$ which corresponds to larger values of the bare $h$.
Thus, for  exchange fields  $h \lesssim \Delta$ the  antiparallel magnetic configuration ($\alpha = \pi$)
of magnetization leads to larger values of the cooling power.  Such  small exchange fields  can be realized in weak ferromagnetic alloys,\cite{small_h} or  in hybrid structures consisting of ferromagnetic
insulators in contact with superconductors.\cite{Tokuyasu1988,Cottet2011} For values  of $h$ larger than $\Delta$,  the parallel
configuration ($\alpha = 0$)  is the one that leads  to larger values of the cooling power.
For values of $h\gg\Delta$ the cooling is almost independent of $\alpha$.

Finally we have analyzed the dependence of the cooling power on the bias voltage, the tunneling parameter and the
temperature.  The optimized values for more efficient cooling are shown in Figs.~\ref{panelv}, \ref{panelW} and \ref{panelT}.


\begin{acknowledgments}
The authors thank E. V. Bezuglyi, S. Kawabata and J. P. Pekola for useful discussions. This work was supported
by  the Spanish Ministry of Economy and Competitiveness under Project FIS2011-28851-C02-02 and the Basque Government under UPV/EHU Project IT-366- 07. The work of A. O. was supported by the CSIC and the European Social Fund under JAE-Predoc program.
A.S.V. acknowledge the hospitality of Donostia International Physics Center (DIPC), during his stay in Spain.
\end{acknowledgments}


\begin{thebibliography}{99}
\bibitem{footnote1}
We neglect the contribution to the Andreev current due to the partial Andreev reflection
at the energies above the superconducting gap. In the case of strong enough tunnel barriers, this contribution
leads to a small correction and can be neglected, see Ref.~\onlinecite{VB} for details.
\end{thebibliography}
\end{document}